# Revisiting Schrödinger's fourth-order, real-valued wave equation and its implications to energy levels

Nicos Makris[1]

**ABSTRACT**

In his seminal part IV, Annalen der Physik Vol 81, 1926 paper, Schrödinger has developed a clear understanding about the wave equation that produces the correct quadratic dispersion relation for matter-waves and he first presents a real-valued wave equation that is 4$^{th}$-order in space and 2$^{nd}$-order in time. In view of the mathematical difficulties associated with the eigenvalue analysis of a 4$^{th}$-order, differential equation in association with the structure of the Hamilton-Jacobi equation, Schrödinger splits the 4$^{th}$-order real operator into the product of two, 2$^{nd}$-order, conjugate complex operators and retains only one of the two complex operators to construct his iconic 2$^{nd}$-order, complex-valued wave equation. In this paper we show that Schrödinger's original 4$^{th}$-order, real-valued wave equation is a stiffer equation that produces higher energy levels than his 2$^{nd}$-order, complex-valued wave equation; that predicted with remarkable success the visible energy levels observed in the visible atomic line-spectra of the chemical elements. Accordingly, the 4$^{th}$-order, real-valued wave equation is too stiff to predict the emitted energy levels from the electrons of the chemical elements; therefore, the paper concludes that Quantum Mechanics can only be described with the less stiff, 2$^{nd}$-order complex-valued wave equation; unless in addition to the emitted visible energy there is also dark energy emitted.

**INTRODUCTION**

During his effort to construct a matter-wave equation that satisfies the quadratic dispersion relation between the angular frequency $\omega$ and the wave number $k$ ($\omega = \frac{\hbar}{2m} k^2$ with $\hbar = \frac{h}{2\pi}$ where $h =$

---

[1] Professor, Dept. of Civil and Environmental Engineering, Southern Methodist Univ, Dallas, TX 75275 (corresponding author). Email: nmakris@smu.edu

$6.62607 \times 10^{-34} \frac{m^2 kg}{s}$ = Planck's constant), Schrödinger in his part IV, 1926 paper [1, 2] reaches a real-valued, fourth-order in space and second-order in time differential equation

$$\left(\frac{1}{m}\nabla^2 - \frac{2}{\hbar^2}V(\boldsymbol{r})\right)^2 \psi(\boldsymbol{r},t) + \frac{4}{\hbar^2}\frac{\partial^2 \psi(\boldsymbol{r},t)}{\partial t^2} = 0 \tag{1}$$

where $m$ is the mass of the elementary, non-relativistic particle and $V(\boldsymbol{r})$ is its potential energy that is only a function of the position $\boldsymbol{r}$. In his 1926 paper [1], Schrödinger explains in his own words: "*Eq. (1) is thus evidently the uniform and general wave equation for the field scalar $\psi$*". He further recognizes that his fourth-order, Eq. (1) resembles the fourth-order, equations of motion that emerge from the theory of elasticity and references the governing equation of a vibrating plate. More precisely, because of the 3-dimensional geometry of atoms, the description of an electron orbiting the atom with Eq. (1) resembles to the equation of motion of a vibrating shell [3-6] which had not been developed at that time.

For standing waves, the spatial and temporal dependence of the matter-wave can be separated

$$\psi(\boldsymbol{r},t) = \psi(\boldsymbol{r})e^{\pm\frac{i}{\hbar}Et} \tag{2}$$

so that

$$\frac{\partial \psi(\boldsymbol{r},t)}{\partial t} = \pm\frac{i}{\hbar}E\psi(\boldsymbol{r},t) \quad and \quad \frac{\partial^2 \psi(\boldsymbol{r},t)}{\partial t^2} = -\frac{E^2}{\hbar^2}\psi(\boldsymbol{r},t) \tag{3}$$

In the interest of simplifying the calculations in the eigenvalue analysis of Eq. (1); in association that $V(\boldsymbol{r})$ does not contain the time, Schrödinger [1,2] substitutes the second of Eq. (3) into Eq. (1) and recasts it in a factored form

$$\left(\frac{1}{m}\nabla^2 - \frac{2}{\hbar^2}V(\boldsymbol{r}) + \frac{2}{\hbar^2}E\right)\left(\frac{1}{m}\nabla^2 - \frac{2}{\hbar^2}V(\boldsymbol{r}) - \frac{2}{\hbar^2}E\right)\psi(\boldsymbol{r}) = 0 \tag{4}$$

He recognizes that Eq. (4) does not vanish by merely setting one of the factors equal to zero given that each factor is an operator. Inspired by the factorized form of his original fourth-order, wave Eq. (1) given by Eq. (4) in association with the structure of the Hamilton-Jacobi equation [7-12], Schrödinger reverts to the first of Eq. (3) to separate the time dependence and settles with his iconic second-order in space and first-order in time complex-valued wave equation [1, 2].

$$i\hbar \frac{\partial \psi(\mathbf{r},t)}{\partial t} = -\frac{\hbar^2}{2m} \nabla^2 \psi(\mathbf{r},t) + V(\mathbf{r})\psi(\mathbf{r},t) \qquad (5)$$

At the end of section §1 of his part IV, 1926 paper [1, 2] Schrödinger indicates that for "*a conservative system, Eq. (5) is essentially equivalent to Eq. (1), as the real operator may be split up into the product of the two conjugate complex operators if V does not contain the time*".

The above equivalence statement is not true, since the fourth-order, real-valued wave equation (1) is a "stiffer" equation than the second-order, complex-valued equation (5), yielding higher eigenvalues and therefore higher energy levels.

The higher energy levels predicted by the stiffer fourth-order, real-valued wave equation (1) than these predicted by the classical second-order, complex-valued, Schrödinger equation (5) are shown in this paper by computing the energy levels of a one-dimensional elementary particle, $\psi(x,t)$, trapped in a square well with finite potential $V$. The paper shows that the one-dimensional version of Schrödinger's original fourth-order, real-valued equation is equivalent to the governing equation of a vibrating flexural-shear beam [13, 14]. By splitting the fourth-order, real-valued operator into the product of two conjugate second-order, complex-valued operators and upon retaining only one of the complex operators, Schrödinger [1, 2] essentially removed from his original fourth-order, Eq. (1) its "flexural stiffness" and left it only with its "shear stiffness".

In view of the many predictions with remarkable precision of Schrödinger's second-order, complex-valued Eq. (5) for the atomic orbitals of the chemical elements [15-19] in association with the higher energy levels predicted from his original fourth-order, real-valued Eq. (1) (therefore, apparently incorrect), this paper offers a straight forward explanation why Quantum Mechanics can only be described with complex-valued functions—a finding that is in agreement with more elaborate recent studies that hinge upon symmetry conditions of real number pairs [20] or involve entangled qubits [21-23]. These recent studies on entangled qubits [21-23] offer the opposite conclusion than the work of McKague et al. [24] which suggests that a real-valued quantum theory can describe a broad range of quantum systems.

This paper shows in a simple, straight-forward manner that Schrödinger's original fourth-order, real-valued wave equation (1), which is the simplest possible real-valued wave equation that satisfies the quadratic dispersion relation, $\omega = \frac{\hbar}{2m} k^2$, is too stiff to predict the visible energy levels that correspond to the visible atomic line spectra of the chemical elements. By splitting the fourth-order, real-valued operator of Eq. (1) into the product of two conjugate second order, complex-valued operators, Schrödinger [1, 2] extracts a more flexible equation than his original 4th-order, real-valued Eq. (1) at the expense of being complex-valued—that is his iconic Eq. (5) which predicted correctly the energy levels of the hydrogen atom; and subsequently made a wealth of fundamental predictions with remarkable precision at the atomic and molecular scale in the century to come [15-19, 25].

The question that deserves an answer is how Schrödinger developed the remarkable intuition to proceed from the onset of his efforts with a complex-valued equation for matter-waves—that is only the one factor of the split 4th order, real-valued equation; which while complex-valued, is flexible enough to predict the correct frequencies manifested in the visible atomic line spectra of the chemical elements in the years to come and abandoned his original fourth-order, real-valued equation that its predictions were apparently never explored.

**THE "FLEXURAL-SHEAR BEAM" EQUATION FOR MATTER-WAVES**

In the interest of illustrating that the fourth-order, real-valued wave equation (1) is a stiffer equation than Schrödinger's second-order, complex-valued Eq. (5), we consider for simplicity a single elementary, non-relativistic practice with mass $m > 0$ in one-dimension moving along the positive direction, $x$, within an energy potential $V(x)$. The total energy of the elementary particle, $E$, is described with its Hamiltonian,

$$E = \mathrm{H}(x, p) = \frac{p^2}{2m} + V(x) \tag{6}$$

where $p = dx/dt$ is the momentum of the elementary particle and $\frac{p^2}{2m} = \frac{1}{2}m\left(\frac{dx}{dt}\right)^2$ represents its kinetic energy. Using Einstein's [26] quantized energy expression, $E = h\nu = \hbar\omega$ and de Broglie's [27] momentum— wavelength relation, $p = h/\lambda = \hbar k$, where $k = 2\pi/\lambda$ is the wave number, the Hamiltonian of the elementary particle given by Eq. (6) assumes the form

$$\hbar\omega = \frac{\hbar^2 k^2}{2m} + V(x) \tag{7}$$

For a particle moving freely in the absence of a potential ($V(x) = 0$), Eq. (7) leads to a quadratic dispersion relation $\omega = \frac{\hbar}{2m} k^2$ for matter-waves as opposed to the linear dissipation relation, $\omega = Ck$, of electromagnetic waves of shear waves in a solid continuum.

The simplest expression for a matter-wave travelling along the positive $x-$ direction is $\psi(x,t) = \psi_0 \, e^{i(kx-\omega t)}$ and upon using that $k = p/\hbar$ and $\omega = E/\hbar$

$$\psi(x,t) = \psi_0 \, e^{\frac{i}{\hbar}(px-Et)} \tag{8}$$

The time-derivative of Eq. (8) gives

$$\frac{\partial \psi(x,t)}{\partial t} = -\frac{i}{\hbar} E \, \psi(x,t) \tag{9}$$

Substitution of the expression for the energy, $E$, given by Eq. (6) into Eq. (9) gives

$$i\hbar \frac{\partial \psi(x,t)}{\partial t} = \left(\frac{p^2}{2m} + V(x)\right) \psi(x,t) \tag{10}$$

The 2$^{nd}$ space-derivative of Eq. (8) gives

$$\frac{\partial^2 \psi(x,t)}{\partial x^2} = -\frac{1}{\hbar^2} p^2 \psi(x,t) \tag{11}$$

and substitution of the quantity $p^2 \psi(x,t)$ from Eq. (11) into Eq. (10) yields the one-dimensional version of the time-dependent Schrödinger equation given by Eq. (5)

$$i\hbar \frac{\partial \psi(x,t)}{\partial t} = -\frac{\hbar^2}{2m} \frac{\partial^2 \psi(x,t)}{\partial x^2} + V(x)\psi(x,t) \tag{12}$$

We now proceed by taking higher derivatives. The time-derivative of Eq. (9) in association with Eq. (8) gives

$$\frac{\partial^2 \psi(x,t)}{\partial t^2} = -\frac{E^2}{\hbar^2} \psi(x,t) \tag{13}$$

whereas by raising the Hamiltonian given by Eq. (6) to the second power gives

$$E^2 = H^2(x,p) = \frac{p^4}{4m^2} + \frac{p^2}{m} V(x) + V^2(x) \tag{14}$$

Substitution of the expression for $E^2$ given by Eq. (14) into Eq. (13) yields

$$\frac{\partial^2 \psi(x,t)}{\partial t^2} = -\frac{1}{\hbar^2}\left(\frac{p^4}{4m^2} + \frac{p^2}{m}V(x) + V^2(x)\right)\psi(x,t) \qquad (15)$$

Upon differentiating of Eq. (11) in space two more times,

$$\frac{\partial^4 \psi(x,t)}{\partial x^4} = \frac{p^4}{\hbar^4}\psi(x,t) \qquad (16)$$

The substitution of the quantity $p^4 \psi(x,t)$ from Eq. (16) and of the quantity $p^2 \psi(x,t)$ from Eq. (11) into Eq. (15) gives

$$-\hbar^2 \frac{\partial^2 \psi(x,t)}{\partial t^2} = \frac{\hbar^4}{4m^2}\frac{\partial^4 \psi(x,t)}{\partial x^4} - \frac{\hbar^2}{m}V(x)\frac{\partial^2 \psi(x,t)}{\partial x^2} + V^2(x)\psi(x,t) \qquad (17)$$

Equation (17) is the one-dimensional version of the real-valued Eq. (1) originally presented by Schrödinger [1, 2] which satisfies the quadratic dispersion relation of matter-waves as dictated by Eq. (7). We coin this time-dependent equation: the "*flexural-shear beam wave equation*" because of the striking similarities with an approximate beam equation that was proposed by Heidebrecht and Stafford Smith [13] to model the dynamics of tall buildings which consist of a strong core-wall that offers flexural resistance acting in parallel with the surrounding framing system of the building that offers shear resistance to lateral loads.

**THE TIME-INDEPENDENT FLEXURAL-SHEAR BEAM EQUATION FOR MATTER-WAVES**

The corresponding time-independent equation for standing waves (mode shapes) of Eq. (17) is derived with the standard method of separation of variables where $\psi(x,t) = \psi(x)f(t)$. Accordingly,

$$\frac{\partial^2 \psi(x,t)}{\partial t^2} = \psi(x)\frac{\mathrm{d}^2 f(t)}{\mathrm{d}t^2} \qquad (18)$$

and

$$\frac{\partial^2 \psi(x,t)}{\partial x^2} = \frac{\mathrm{d}^2 \psi(x)}{\mathrm{d}x^2}f(t); \qquad \frac{\partial^4 \psi(x,t)}{\partial x^4} = \frac{\mathrm{d}^4 \psi(x)}{\mathrm{d}x^4}f(t) \qquad (19)$$

Substitution of the expressions for the partial derivatives given by Eqs. (18) and (19) into Eq. (17) and upon dividing with $\psi(x)f(t)$ gives

$$-m\frac{1}{\psi(t)}\frac{d^2 f(t)}{dt^2} = \frac{\hbar^2}{4m}\frac{1}{\psi(x)}\frac{d^4\psi(x)}{dx^4} - \frac{V(x)}{\psi(x)}\frac{d^2\psi(x)}{dx^2} + \frac{m}{\hbar^2}V^2(x) \qquad (20)$$

The left hand-side of Eq. (20) is a function of time alone; whereas, the right-hand side is a function of space alone. Accordingly,

$$-m\frac{1}{f(t)}\frac{d^2 f(t)}{dt^2} = K \qquad (21)$$

where $K$ is a spring constant with units $[M][T]^{-2}$.

Accordingly, Eq. (21) is the equation of motion of a harmonic oscillator with a real-valued solution

$$f(t) = A \sin \omega t + B \cos \omega t \qquad (22)$$

where $\omega = \sqrt{K/m}$ is the natural frequency of the harmonic oscillator.

Returning to Eq. (20), its right-hand side is also equal to the spring constant $K = m\omega^2$.

$$\frac{\hbar^2}{4m}\frac{1}{\psi(x)}\frac{d^4\psi(x)}{dx^4} - \frac{V(x)}{\psi(x)}\frac{d^2\psi(x)}{dx^2} + \frac{m}{\hbar^2}V^2(x) = m\omega^2 \qquad (23)$$

Multiplication of Eq. (23) with $\hbar^2\psi(x)/m$ yields the time-independent flexural-shear beam equation for matter-waves

$$\frac{\hbar^4}{4m^2}\frac{d^4\psi(x)}{dx^4} - \frac{\hbar^2}{m}V(x)\frac{d^2\psi(x)}{dx^2} + V^2(x)\psi(x) = E^2\psi(x) \qquad (24)$$

where $E = \hbar\omega$ is the quantized energy of the elementary particle. The solution of Eq. (24) yields the eigenvalues and eigenmodes. From the first space derivative of Eq. (8), $\partial\psi(x,t)/\partial x = (i/\hbar)p\psi(x,t)$, we define the standard momentum operator, $\hat{p} = -i\hbar\frac{\partial}{\partial x}$. Accordingly, from Eq. (11), the momentum square operator $\hat{p}^2 = -\hbar^2\frac{\partial^2}{\partial x^2}$ and from Eq. (6), the Hamiltonian operator is:

$$\widehat{H} = \frac{\hat{p}^2}{2m} + V(x) = -\frac{\hbar^2}{2m}\frac{\partial^2}{\partial x^2} + V(x) \qquad (25)$$

From Eq. (25), the Hamiltonian square operator $\widehat{H}^2$ assumes the expression

$$\widehat{H}^2 = \frac{h^4}{4m^2}\frac{\partial^4}{\partial x^4} - \frac{\hbar^2}{m}V(x)\frac{\partial^2}{\partial x^2} + V^2(x) \qquad (26)$$

Accordingly, by employing the Hamiltonian square operator $\widehat{H}^2$ defined by Eq. (26), the time-independent flexural-shear beam equation (24) can be expressed in the compact from

$$\widehat{H}^2\psi(x) = E^2\psi(x) \qquad (27)$$

It is the Hamiltonian square operator $\widehat{H}^2$ that renders Eq. (27) stiffer than the classical time-independent Schrödinger equation $\widehat{H}\psi(x) = E\psi(x)$ that was depleted from its original flexural stiffness [1, 2].

**ELEMENTARY PARTICLE TRAPPED IN A FINITE POTENTIAL SQUARE WELL WITH STRENGTH $V > 0$**

Given that both the 4$^{th}$-order, real-valued flexural-shear beam Eq. (17) and the 2$^{nd}$-order, complex-valued Schrödinger Eq. (12) satisfy the quadratic dispersion relation offered by Eq. (7) as dictated by the Hamiltonian; we proceed by comparing the prediction of these two equations in an effort to show that Schrödinger's original, fourth-order, real-valued Eq. (1) is a stiffer differential equation than his second-order, complex-valued Eq. (5) or Eq. (12) in one dimension. The quadratic Hamiltonian operator appearing in the flexural-shear beam Eq. (27) leads to elaborate calculations even for simple cases; therefore, we select as a test case the response analysis of an elementary, particle with mass $m$ trapped in a square potential well with finite potential $V$ and width $2L$. Accordingly, the potential at the bottom of the well is zero as shown in Fig. 1. This simple, one dimensional idealization has been employed to determine the wavelengths for color-center absorption [28].

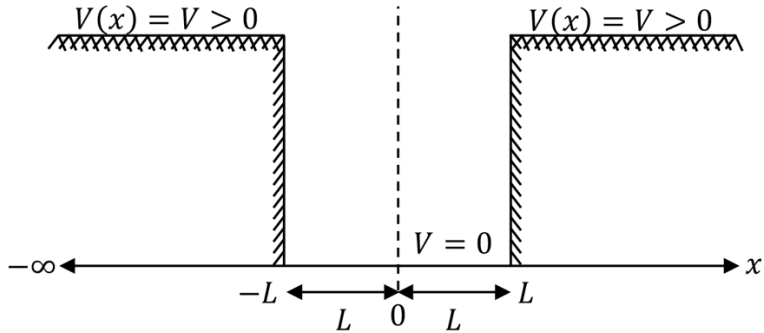

**Fig. 1.** The finite potential square well with constant strength $V$ outside the well with width $2L$.

For the case where the elementary particle happens to be outside the well ($|x| \geq L$), $V(x) = V > 0$, and Eq. (24) gives

$$\frac{d^4\psi(x)}{dx^4} - \frac{4m}{\hbar^2}V\frac{d^2\psi(x)}{dx^2} + \frac{4m^2}{\hbar^4}(V^2 - E^2)\psi(x) = 0 \quad (28)$$

The solutions of the homogeneous Eq. (28) are expected to be of the form $\psi(x) = e^{\beta x}$ and Eq. (28) yields the following characteristic equation

$$\beta^4 - \frac{4m}{\hbar^2}V\beta^2 + \frac{4m^2}{\hbar^4}(V^2 - E^2) = 0 \quad (29)$$

where $V > E > 0$. The four roots of the characteristic Eq. (29) are

$$\beta_1 = \frac{1}{\hbar}\sqrt{2m(V+E)} > 0, \qquad \beta_2 = -\frac{1}{\hbar}\sqrt{2m(V+E)} = -\beta_1 \quad (30)$$

$$\beta_3 = \frac{1}{\hbar}\sqrt{2m(V-E)} > 0, \qquad \beta_4 = -\frac{1}{\hbar}\sqrt{2m(V-E)} = -\beta_3 \quad (31)$$

Accordingly, for the case $|x| \geq L$ where $V(x) = V > E > 0$ the solution for $\psi(x)$ is

$$\psi(x) = A_1 e^{\beta_1 x} + A_2 e^{-\beta_2 x} + A_3 e^{\beta_3 x} + A_4 e^{-\beta_4 x} \quad (32)$$

For the case where the elementary particle is within the potential well ($|x| \leq L$), $V(x) = 0$ and Eq. (24) gives

$$\frac{d^4\psi(x)}{dx^4} - \frac{4m^2}{\hbar^4}E^2\psi(x) = 0 \quad (33)$$

By setting $(4m^2/\hbar^4)E^2 = k^4$, Eq. (33) assumes the form

$$\frac{d^4\psi(x)}{dx^4} - k^4\psi(x) = 0 \quad (34)$$

Eq. (34) has a real-valued solution [29, 30].

$$\psi(x) = C_1 \sin(kx) + C_2 \cos(kx) + C_3 \sinh(kx) + C_4 \cosh(kx) \quad (35)$$

where $k = (1/\hbar)\sqrt{2mE}$ is a positive, real wave number. In this case $(x \leq |L|)$, $V(x) = 0$ and from Eq. (6), $E = p^2/(2m)$, therefore, the wave number $k = (1/\hbar)\sqrt{2mE}$ appearing in Eq. (35)

is $k = (1/\hbar)\sqrt{2mp^2/(2m)} = p/\hbar$ is the de Broglie wave number. This supports, the choice for the same symbol, $k$.

It is worth nothing that Eq. (34) is the equation of motion of a vibrating flexural beam with distributed mass per length $\bar{m}$ with units $[M][L]^{-1}$, Young's modulus of elasticity, $Y$ with units $[M][L][T]^{-2}$ ($force/area$) and moment of cross-sectional area, $I$ with units $[L]^4$. For a vibrating flexural beam $k^4 = \bar{m}\omega^2/YI$ and upon using $E = \hbar\omega$ and cancelling the angular frequency $\omega$, we obtain the analogy $YI/\bar{m} \rightarrow (\hbar/2m)^2$, both having units of $[L]^4[T]^{-2}$.

**Continuity of solutions**

*Case 1: $x \leq -L$ where $V(x) = V$ and $V - E > 0$. Bound states.*

For this case where $x \leq -L$, the solution $\psi(x)$ given by Eq. (32) remains finite when $A_2 = A_4 = 0$. Consequently, for this case

$$\psi(x) = A_1 e^{\beta_1 x} + A_3 e^{\beta_3 x} \qquad for \ x \leq -L \tag{36}$$

in which $\beta_1$ and $\beta_3$ are real-valued and given by Eqs. (30) and (31).

*Case 2: $-L \leq x < L$ where $V(x) = 0$.*

For this case $\psi(x)$ is given by Eq. (35).

*Case 3: $x \geq L$ where $V(x) = V$ and $V - E > 0$. Bound states.*

For this case where $x > L$, the solution $\psi(x)$ given by Eq. (32) remains finite when $A_1 = A_3 = 0$. Consequently, for this case

$$\psi(x) = A_2 e^{-\beta_1 x} + A_4 e^{-\beta_3 x} \qquad for \ x > L \tag{37}$$

in which $\beta_1$ and $\beta_3$ are real-valued and given by Eqs. (30) and (31).

The solution of the wave equation $\psi(x)$ has to be continuous over the entire domain $-\infty < x < \infty$. Accordingly, at $x = -L$, Eq. (36) from the left and Eq. (35) from the right need to satisfy the following continuity equations:

$$\psi(-L^-) = \psi(-L^+), \qquad \frac{d\psi(-L^-)}{dx} = \frac{d\psi(-L^+)}{dx} \tag{38a}$$

$$\frac{d^2\psi(-L^-)}{dx^2} = \frac{d^2\psi(-L^+)}{d^2 x}, \quad \frac{d^3\psi(-L^-)}{d^3 x} = \frac{d^3\psi(-L^+)}{d^3 x} \tag{38b}$$

Similarly, at $x = L$, Eq. (35) from the left and Eq. (37) from the right need to satisfy the following continuity equations.

$$\psi(L^-) = \psi(L^+), \quad \frac{d\psi(L^-)}{dx} = \frac{d\psi(L^+)}{dx} \tag{39a}$$

$$\frac{d^2\psi(L^-)}{dx^2} = \frac{d^2\psi(L^+)}{d^2 x}, \quad \frac{d^3\psi(L^-)}{d^3 x} = \frac{d^3\psi(L^+)}{d^3 x} \tag{39b}$$

The eight continuity equations given by Eqs. (38) and (39) form a homogeneous system of 8 equations which yields the eigenvalues $z_n = k_n L$ and eigenfunctions (mode shapes) $\psi_n(x)$ of the wavefunction $\psi(x)$.

**Eigenvalue analysis**

The wavenumbers $\beta_1$ and $\beta_3$ given by Eq. (30) and (31) can be expressed as

$$\beta_1 = \sqrt{\frac{2mV}{\hbar^2} + \frac{2mE}{\hbar^2}} = \sqrt{b^2 + k^2} \tag{40}$$

$$\beta_3 = \sqrt{\frac{2mV}{\hbar^2} - \frac{2mE}{\hbar^2}} = \sqrt{b^2 - k^2} \tag{41}$$

where $b = (1/\hbar)\sqrt{2mV}$ is a positive number and $k = (1/\hbar)\sqrt{2mE} = 2\pi/\lambda = p/\hbar$ is the wavenumber of the solution of $\psi(x)$ when $-L \leq x \leq L$ given by Eq. (35).

The homogeneous system of eight equations that is generated by the eight continuity Eqs. (38) and (39) can be decomposed in four equations that produce the even eigenfunctions $\psi_n^e(x)$ and four equations produce the odd eigenfunctions $\psi_n^o(x)$. The homogeneous system that produces the even eigenfunctions is

$$\begin{bmatrix} \cos(z) & \cosh(z) & -e^{-\sqrt{b^2L^2+z^2}} & -e^{-\sqrt{b^2L^2-z^2}} \\ -z\sin(z) & z\sinh(z) & \sqrt{b^2L^2+z^2}\,e^{-\sqrt{b^2L^2+z^2}} & \sqrt{b^2L^2-z^2}\,e^{-\sqrt{b^2L^2-z^2}} \\ -z^2\cos(z) & z^2\cosh(z) & -(b^2L^2+z^2)\,e^{-\sqrt{b^2L^2+z^2}} & -(b^2L^2-z^2)\,e^{-\sqrt{b^2L^2-z^2}} \\ z^3\sin(z) & z^3\sinh(z) & (b^2L^2+z^2)^{3/2}\,e^{-\sqrt{b^2L^2+z^2}} & (b^2L^2-z^2)^{3/2}\,e^{-\sqrt{b^2L^2-z^2}} \end{bmatrix} \begin{Bmatrix} C_2 \\ C_4 \\ A_2 \\ A_4 \end{Bmatrix} = 0 \quad (42)$$

where $bL = (L/\hbar)\sqrt{2mV}$ is a dimensionless positive real number that expresses the strength of the potential well and $z = kL = (L/\hbar)\sqrt{2mE}$ are the eigenvalues of the even eigenfunctions to be determined. The eigenvalues $z_n$ depend on the dimensionless product $bL$ rather than on the individual values of $b$ and $L$ and they are calculated by setting the determinant of the $4 \times 4$ matrix appearing on the left of Eq. (42) equal to zero. As an example, for $bL = 10$ the characteristic equation of the homogeneous system given by Eq. (42) yields four real roots (eigenvalues, $n \in \{1,2,3,4\}$) for $z_n = (L/\hbar)\sqrt{2mE_n} = 1.9747, 4.6204, 7.2901$, and $9.7999$. For larger value of $bL$ (deeper and wider potential well) the number of real eigenvalues increases given that the unknown eigenvalue $z$ needs to remain smaller than $bL$ for the radical $\sqrt{b^2L^2 - z^2}$ of the last column of the matrix appearing in Eq. (42) to remain positive.

Similarly, the homogenous system as results from the continuity equations that produces the odd eigenfunctions is

$$\begin{bmatrix} \sin(z) & \sinh(z) & -e^{-\sqrt{b^2L^2+z^2}} & -e^{-\sqrt{b^2L^2-z^2}} \\ z\cos(z) & z\cosh(z) & \sqrt{b^2L^2+z^2}\,e^{-\sqrt{b^2L^2+z^2}} & \sqrt{b^2L^2-z^2}\,e^{-\sqrt{b^2L^2-z^2}} \\ -z^2\sin(z) & z^2\sinh(z) & -(b^2L^2+z^2)\,e^{-\sqrt{b^2L^2+z^2}} & -(b^2L^2-z^2)\,e^{-\sqrt{b^2L^2-z^2}} \\ -z^3\cos(z) & z^3\cosh(z) & (b^2L^2+z^2)^{3/2}\,e^{-\sqrt{b^2L^2+z^2}} & (b^2L^2-z^2)^{3/2}\,e^{-\sqrt{b^2L^2-z^2}} \end{bmatrix} \begin{Bmatrix} C_1 \\ C_3 \\ A_2 \\ A_4 \end{Bmatrix} = 0 \quad (43)$$

The finite eigenvalues $z_n = (L/\hbar)\sqrt{2mE_n}$ that corresponds to the odd eigenfunctions are computed by setting the determinant of the $4 \times 4$ matrix appearing on the left of Eq. (43) equal to zero. As an example, for $bL = 10$ the characteristic equation of the homogeneous system given by Eq. (43) yields three real roots (eigenvalues, $n \in \{1,2,3\}$) for $z_n = (L/\hbar)\sqrt{2mE_n} = 3.2887, 5.9574$, and $8.5976$. For larger value of $bL$ (deeper and wider potential well) the number of real roots of the characteristic equation (eigenvalues) increases as long as $z < bL$ so that the radical $\sqrt{b^2L^2 - z^2}$ appearing in the last column of the $4 \times 4$ matrix Eq. (43) remains real.

# COMPARISON OF THE EIGENVALUES PREDICTED FROM THE 4$^{TH}$-ORDER FLEXURAL-SHEAR BEAM EQUATION AND FROM THE CLASSICAL 2$^{ND}$-ORDER SCHRÖDINGER EQUATION

For any given value of the strength of the square potential well, $bL$ the resulting eigenvalues of the fourth-order, flexural-shear beam Eq. (24) or Eq. (27), $z_n = (L/\hbar)\sqrt{2mE_n}$, yield the admissible energy levels of the elementary particle in the finite square potential well, $E_n = (z_n^2 \hbar^2)/(2mL^2)$. Clearly, the predicted energy levels, $E_n$, are different than the corresponding energy levels, $E_n$, predicted from the solution of the second-order, time-independent Schrödinger equation.

The predicted eigenvalues, $z_n = (L/\hbar)\sqrt{2mE_n}$ of an elementary particle in a finite square potential well with the second-order, Schrödinger equation are the roots of the transcendental Eqs. (44) and (45) [31]

$$\tan(z) = \sqrt{\frac{b^2 L^2}{z^2} - 1} \qquad for\ even\ eigenfuctions \qquad (44)$$

and

$$\cot(z) = -\sqrt{\frac{b^2 L^2}{z^2} - 1} \qquad for\ odd\ eigenfuctions \qquad (45)$$

where $b = (1/\hbar)\sqrt{2mV}$ as in the previous analysis.

As an example for $bL = 10$, Eq. (44) yields four real roots (eigenvalues of the even eigenfunctions, $n \in \{1, 2, 3, 4\}$) for $z_n = (L/\hbar)\sqrt{2mE_n} = 1.4276, 4.2711, 7.0689, 9.6789$; and Eq. (45) yields three real roots (eigenvalues of the odd eigenfunctions, $n \in \{1, 2, 3\}$) for $z_n = (L/\hbar)\sqrt{2mE_n} = 2.8523, 5.6792$, and $8.4232$.

Table 1 compares the predicted eigenvalues for a non-relativistic particle in a finite square potential well with potential $V$ from the fourth-order, flexural-shear beam wave equation and the second-order, Schrödinger wave equation for $bL = 10$ and 30. Table 1 also shows the limiting eigenvalues for a particle trapped in an infinitely deep potential well ($V = \infty$) as they result from the second-order, Schrödinger equation, $z_n = (L/\hbar)\sqrt{2mE_n} = n\pi/2$ [31] and from the fourth-order, flexural-shear beam equation which are the solution of the characteristic equation $\cos(2kL)\cosh(2kL) = 1$ as shown in the sequel.

**Table 1**: The seven eigenvalues (energy levels) $z_n = \frac{L}{\hbar}\sqrt{2mE_n}$ for a particle in a finite potential well with strength $bL = \frac{L}{\hbar}\sqrt{2mV} = 10$, when described with the fourth-order, flexural-shear beam wave equation and with the classical second-order, Schrödinger wave equation, together with the first 9 corresponding eigenvalues when $bL = 30$ and $\infty$.

| No of Eigenvalue $z_n = \frac{L}{\hbar}\sqrt{2mE_n}$ | 4$^{\text{th}}$-order Flexural-Shear Beam Eq. | | | 2$^{\text{nd}}$-order Schrödinger Eq. | | |
|---|---|---|---|---|---|---|
| | $bL = \frac{L}{\hbar}\sqrt{2mV}$ | | $\frac{\cos(2z)}{1} = \frac{1}{\cosh(2z)}$ | $bL = \frac{L}{\hbar}\sqrt{2mV}$ | | $z_n = \frac{L}{\hbar}\sqrt{2mE_n} = \frac{n\pi}{2}$ |
| | $bL = 10$ | $bL = 30$ | $bL = \infty$ | $bL = 10$ | $bL = 30$ | $bL = \infty$ |
| $n = 1$ | 1.974707 | 2.217448 | 2.365020 | 1.427552 | 1.520104 | $\frac{\pi}{2} = 1.570796$ |
| $n = 2$ | 3.288725 | 3.682318 | 3.926602 | 2.852342 | 3.040082 | $\pi = 3.141593$ |
| $n = 3$ | 4.620365 | 5.157210 | 5.497804 | 4.271095 | 4.559804 | $\frac{3\pi}{2} = 4.712389$ |
| $n = 4$ | 5.957359 | 6.633016 | 7.068583 | 5.679208 | 6.079134 | $2\pi = 6.283185$ |
| $n = 5$ | 7.290139 | 8.113046 | 8.639380 | 7.068891 | 7.597928 | $\frac{5\pi}{2} = 7.853982$ |
| $n = 6$ | 8.597635 | 9.589274 | 10.210176 | 8.423204 | 9.116028 | $3\pi = 9.424778$ |
| $n = 7$ | 9.799891 | 11.06978 | 11.780972 | 9.678884 | 10.633257 | $\frac{7\pi}{2} = 10.995574$ |
| $n = 8$ | – – – | 12.55174 | 13.351769 | – – – | 12.149413 | $4\pi = 12.566371$ |
| $n = 9$ | – – – | 14.03491 | 14.922565 | – – – | 13.664261 | $\frac{9\pi}{2} = 14.137167$ |
| ⋮ | – – – | ⋮ | ⋮ | – – – | ⋮ | ⋮ |

Table 1 reveals that when $bL = 10$ all seven eigenvalues that result from the fourth-order, flexural-shear beam equation are larger than the corresponding seven eigenvalues that result from the classical second-order, Schrödinger equation. The same is true for the case when $bL = 30$. Consequently, this analysis shows that the 4$^{\text{th}}$-order, real-valued flexural-shear beam equation for matter-waves given by Eq. (17) is a stiffer equation than the classical 2$^{\text{nd}}$-order, complex-valued Schrödinger equation given by Eq. (13). Therefore, Schrödinger's equivalence statement that Eq.

(5) (which is Eq. (4″) in his 1926 paper [1]); and Eq. (1) (which is Eq. (4) in his 1926 paper [1]) are equivalent, is not true.

Furthermore, Table 1 reveals that when $bL = 10$, the first two eigenvalues $z_1 = 1.9747$ and $z_2 = 3.2887$ that result from the fourth-order, flexural-shear beam equation are even larger than the first two eigenvalues $z_1 = \pi/2$, and $z_2 = \pi$ that result from the classical second-order, Schrödinger equation at the limiting case when the strength of the potential well is infinite ($bL = \frac{t}{\hbar}\sqrt{2mV} = \infty$) [31]. This pattern where the eigenvalues predicted from the fourth-order, flexural-shear beam equation when trapped in a finite potential well exceed the eigenvalues predicted by the second-order, Schrödinger equation when the particle is trapped in an infinite potential well becomes more dominant as the strength, $bL$ of the finite potential well increases. For instance, when $bL = 30$, the first seven eigenvalues that result from the fourth-order, flexural-shear beam equation are larger than the first seven eigenvalues that result from the classical second-order, Schrödinger equation at the limiting case of an infinitely strong potential well. Accordingly, there is a need to calculate the energy levels of an elementary particle trapped in an infinitely strong potential well ($bL = \infty$) when described with the fourth-order, flexural-shear beam wave equation (24) or (27).

The wavefunctions (eigenmodes) associated with the energy levels (eigenvalues) appearing in Table 1 for the situation where the elementary particle is described with the fourth-order, flexural-shear beam wavefunction are offered by Eq. (35) for $0 \leq |x| \leq L$ and by Eq. (37) for $x \geq L$. Accordingly, the even eigenfunctions, ($n \in \{1, 3, 5, \dots\}$) are given by

$$\psi_n^e(x) = C_2 \cos\left(z_n \frac{x}{L}\right) + C_4 \cosh\left(z_n \frac{x}{L}\right) \qquad for\ 0 \leq |x| < L \qquad (46)$$

and

$$\psi_n^e(x) = A_2 e^{-\sqrt{b^2L^2 + z_n^2}\,\frac{x}{L}} + A_4 e^{-\sqrt{b^2L^2 - z_n^2}\,\frac{x}{L}} \qquad for\ L < x; \qquad (47)$$

whereas, the odd eigenfunctions, ($n \in \{2, 4, 6, \dots\}$) are given by

$$\psi_n^o(x) = C_1 \sin\left(z_n \frac{x}{L}\right) + C_3 \sinh\left(z_n \frac{x}{L}\right) \qquad for\ 0 \leq |x| < L \qquad (48)$$

and $\psi_n^o(x)$ is given again by Eq. (47) for $L < x$.

The coefficients $C_2$, $C_4$, $A_2$ and $A_4$ appearing in Eqs. (46) and (47) are obtained upon solving the homogeneous system of equations given by the matrix Eq. (42); whereas, the coefficients $C_1$, $C_2$, $A_2$, and $A_4$ appearing in Eqs. (48) and (49) are obtained upon solving the homogenous system of equations given by the matrix Eq. (43). When solving the homogeneous system of equations, one of the four coefficients is assigned an arbitrary value and the other three coefficients are calculated in proportion to the arbitrary assigned value of the first coefficient since the eigenfunctions $\psi_n^e(x)$ and $\psi_n^o(x)$ are eigenmodes of arbitrary amplitude which subsequently can be normalized according to some normalizing rule such as $\int_{-\infty}^{\infty}|\psi(x)|^2 dx = \int_{-\infty}^{\infty}\psi^2(x)dx = 1$, where the norm, | |, has been dropped from the second integral given that in this case $\psi(x)$ is a real-valued function.

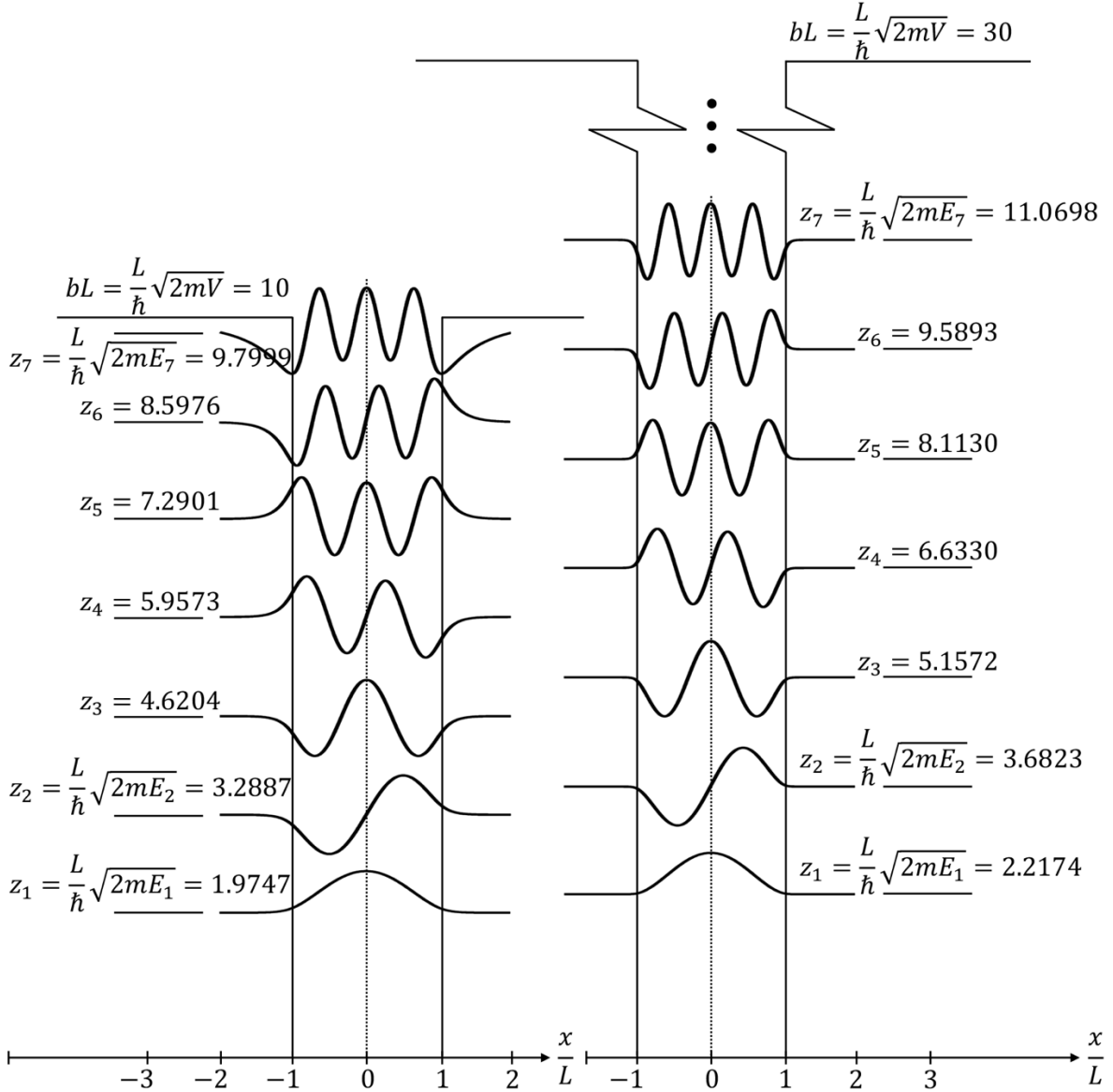

**Fig. 2.** The seven eigenfunctions $\psi_n(x),\ n \in \{1, 2, \ldots, 7\}$, of an elementary particle described with the 4$^{th}$-order, flexural-shear beam equation (24) or (27) when trapped in a potential well with finite strength $bL = (L/\hbar)\sqrt{2mV} = 10$ which manifest at the energy levels $E_n = \frac{z_n^2 \hbar^2}{2mL^2}$ (left); together, with the corresponding first seven eigenfunctions $\psi_n(x)$ when the elementary particle is trapped in a potential well with finite strength $bL = (L/\hbar)\sqrt{2mV} = 30$ (right).

Figure (2) (left) plots the seven eigenfunctions $\psi_n(x)$, $n \in \{1, 2, ..., 7\}$ of an elementary, non-relativistic particle described with the fourth-order, flexural-shear beam Eq. (24) or (27) when trapped in a potential well with finite strength, $bL = \frac{t}{\hbar}\sqrt{2mV} = 10$ which manifest at the energy levels $E_n = \frac{z_n^2 \hbar^2}{2mL^2}$. The eigenvalues $z_n$ are listed in Table 1. Fig. 2(right) plots the corresponding first seven wavefunctions $\psi_n(x)$ (they are 19 wavefunctions in total) when the elementary particle is trapped in a potential well with finite strength $bL = 30$.

## EIGENVALUES OF THE 4$^{TH}$-ORDER MATTER-WAVE EQUATION OF AN ELEMENTARY PARTICLE TRAPPED IN AN INFINITE-POTENTIAL SQUARE WELL

Figure 2 reveals that as the strength of the finite potential well increases, the eigenfunctions $\psi_n(x)$ that result from the solution of the 4$^{th}$-order, wave equation (24) or (27) meet the walls of the square potential well at a decreasing slope which eventually tends to zero, $\frac{d\psi}{dx}(x = -L) = \frac{d\psi}{dx}(x = L) = 0$, as the strength of the potential well, $bL$, tends to infinity.

These zero-slope boundary conditions of the eigenmodes of the trapped particle at the walls of the infinitely strong potential well are drastically different than the finite-slope boundary conditions of the eigenmodes of the trapped particle when described with the 2$^{nd}$-order, Schrödinger equation ($\psi_n(x) = \sqrt{\frac{2}{a}}\sin\left(\frac{n\pi}{a}x\right)$ with $0 < x < a = 2L$) [31]. These fixed-end (zero-slope) boundary conditions (clamped eigenmodes) is another proof that the 4$^{th}$-order, real-valued Eq. (1) originally proposed by Schrödinger [1, 2] is a stiffer equation than his classical 2$^{nd}$-order, complex-valued Eq. (5).

The eigenfunctions of the particle trapped in an infinitely strong potential well when described with the 4$^{th}$-order, flexural-shear beam wave equation (24) are given by Eq. (35), and the integration constants $C_1$, $C_2$, $C_3$, and $C_4$ are derived by enforcing the boundary conditions

$$\psi(-L) = \psi(L) = 0 \tag{49a}$$

and

$$\frac{d\psi(-L)}{dx} = \frac{d\psi(L)}{dx} = 0 \tag{49b}$$

This homogeneous system of four equations results to the transcendental characteristic equation

$$\cos(2kL)\cosh(2kL) = 1 \qquad (50)$$

The roots of Eq. (50), $z_n = k_n L = (L/\hbar)\sqrt{2mE_n}$ are the eigenvalues of the fixed-end eigenmodes appearing in Table 1 under $bL = \infty$.

**DISCUSSION AND SUMMARY**

In this paper we show that Schrödinger's original 4$^{th}$-order, real-valued Eq. (1) for matter-waves is a stiffer description (higher energy levels) of the behavior of elementary particles than the description offered from his classical, 2$^{nd}$-order, complex-valued Eq. (5). Given the remarkable predictions of the complex-valued Eq. (5) for the visible energy levels of the chemical elements as manifested from their visible atomic line-spectra [15-19, 25], in association that his original 4$^{th}$-order, real-valued equation predicts invariably higher-energy levels (therefore, apparently incorrect) this paper shows that Quantum Mechanics can only be described with the less stiff, complex-valued wave equation (5). This finding is in agreement with more elaborate recent studies that hinge upon symmetry conditions of real number pairs [20] or involve entangled qubits [21-23].

At the same time, the paper brings forward that Schrödinger's 2$^{nd}$-order, complex-valued equation was extracted from his original 4$^{th}$-order, real-valued equation by splitting the 4$^{th}$-order, real operator in Eq. (1) into the product of two conjugate complex operators and subsequently retaining only one of the two complex 2$^{nd}$-order, operators—a rather disruptive mathematical intervention that removed the "flexural stiffness" from his original 4$^{th}$-order, real-valued equation.

This disruptive mathematical intervention that alters the physics of his original Eq. (1) motivates the conjecture that perhaps the visible energy levels of the chemical elements as manifested from the visible atomic line-spectra is only a fraction of the total emitted energy by the atoms and molecules. The conjecture advanced herein is that perhaps Schrödinger's original 4$^{th}$-order, real-valued Eq. (1) is the correct equation that predicts the total emitted energy (visible and dark) and the excess of energy above the visible energy predicted with Schrödinger's 2$^{nd}$-order complex-valued Eq. (1) is merely dark-energy that is not visible on the atomic line spectra of the chemical elements. In this event,

$$E_{real}^{4^{th}\ order} = E_{visible}^{2^{nd}\ order} + E_{dark} \qquad (51)$$

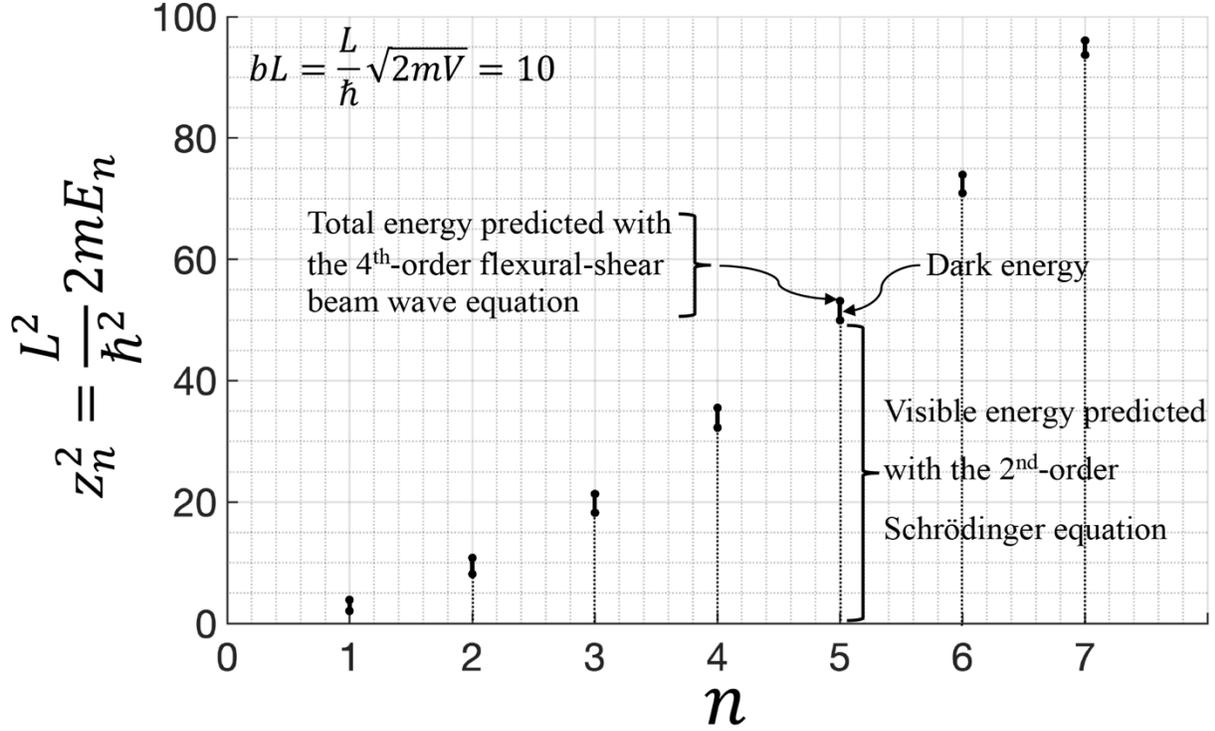

**Fig. 3.** Normalized energy levels, $z_n^2 = (L^2/\hbar^2)2mE_n$ of an elementary particle trapped in a finite potential well with strength $bL = (L/\hbar)\sqrt{2mV} = 10$ predicted with the 4th-order, flexural-shear beam equation (24) and the 2nd-order, classical Schrödinger's equation.

As an example, Fig. 3 plots the normalized energy levels $z_n^2 = (L^2/\hbar^2)2mE_n$ of an elementary particle trapped in a finite-potential square well with strength, $bL = (L/\hbar)\sqrt{2mV} = 10$ as predicted by the 4th-order, flexural-shear beam equation (24) or (27) and the 2nd-order, classical Schrödinger equation. The numerical values of the seven eigenvalues $z_n$ are offered in Table 1. The proportion of the dark energy (dark bars) to the visible energy (dotted lines) shown in Fig. 3 for the one-dimensional idealization of an electron trapped in a finite potential square well is much smaller when compared to the current estimate that roughly 68% of the universe is dark energy. In a realistic 3-dimensional analysis for the energies emitted by the electrons of the chemical elements, the differences between the energy levels predicted from the 3-dimentional "shell" Eq. (1) and the 3-dimentional "membrane" Eq. (5) are expected to be much higher than the difference from their one-dimensional versions; and therefore, the predictions of Schrödinger's original 4th-order Eq. (1) perhaps deserve to be further investigated.